\documentclass[a4paper, 11pt]{article}
\usepackage{amssymb}
\usepackage{enumerate}
\usepackage{natbib}
\usepackage{amsthm}

\usepackage{syntonly,color,subfig,caption}
\usepackage{graphicx}

\usepackage[dvipsnames]{xcolor}
\usepackage{verbatim}
\usepackage[titletoc,title]{appendix}
\usepackage{xcolor}
\usepackage[normalem]{ulem}

\usepackage{amsmath}
\usepackage{mathptmx}

\usepackage{soul}
\definecolor{lightgreen}{rgb}{.85,1,.95}
\sethlcolor{lightgreen}

\definecolor{myrev}{rgb}{0.2,0.21,0.63}
\sethlcolor{myrev}

\newcommand{\revision}[1]{\textcolor{myrev}{#1}} 
\renewcommand\revision[1]{#1}

\definecolor {navy}{HTML}{0080AC} 
\usepackage{hyperref}
\hypersetup{
     colorlinks   = true,
linkcolor = navy, 
anchorcolor = blue,
citecolor =navy,
filecolor =cyan,
menucolor =red,
runcolor =cyan,
urlcolor =navy
}

\usepackage[onehalfspacing]{setspace}

\theoremstyle {plain}     

\newtheorem {proposition}{Proposition}
\newtheorem {lemma}{Lemma}

\theoremstyle {definition}     

\newtheorem {example}{Example}

\newtheorem {fact}{Fact}
\theoremstyle {remark} 
\input{tcilatex}

\begin{document}

\title{Perfect bidder collusion through bribe and request\thanks{ We are grateful to the editors in charge and an anonymous reviewer for their insightful comments and suggestions, which greatly improved the quality of our paper. Any remaining errors are our own.} } 
\author{Jingfeng Lu\thanks{Department of Economics, National University of Singapore. Email: \href{mailto:ecsljf@nus.edu.sg}{ecsljf@nus.edu.sg}.}\quad Zongwei Lu \thanks{School of Economics, Shandong University. Email: \href{mailto:zongwei.lu@sdu.edu.cn}{zongwei.lu@sdu.edu.cn}. } \quad Christian Riis\thanks{Department of Economics, BI Norwegian Business School.
Email: \href{mailto:christian.riis@bi.no}{christian.riis@bi.no}.
}
}
\date{\today} \maketitle

\begin{abstract} We study collusion in a second-price auction with two
bidders in a dynamic environment. One bidder can make a take-it-or-leave-it
collusion proposal, which consists of both an offer and a
request of bribes, to the opponent. We show that there always exists
a robust equilibrium in which the collusion success probability is one. 
\revision{In the equilibrium, for each type
of  initiator the expected payoff is generally higher than the
counterpart in any robust equilibria of the single-option model (\citet{ES2004}) and any other separating equilibria in our model.
}

\vspace{.5cm}

\vspace{.5cm}

\end{abstract}
\clearpage

\section{Introduction}

Under standard assumptions, auctions are a simple yet effective economic institution that allows the owner of a scarce resource to extract economic rents from buyers. Collusion among bidders, on the other hand, can ruin the rents and hence should be one of the main issues that are always kept in
the minds of auction designers with a revenue-maximizing objective. In the literature, collusion in auctions is typically modelled as a static game in which bidders have already agreed to form a cartel and the mission of the cartel is to select a representative bidder (to win the object at a low price in the auction) and decide on the side-payments (e.g., \citet{Graham1987}, \citet{Mailath1991}, \cite{marshall2007},  and \citet{mcafee1992}). Although collusion can be implemented through a revelation game, the working of such a cartel typically requires the aid of an  incentiveless third party. Furthermore, the static model misses a realistic and important scenario in which a particular bidder has strong bargaining power and thus is interested in monetizing it through some bargaining process with other bidders.

In a pioneering work, \citet{ES2004}  (hereafter ES) consider a dynamic model of bidder collusion in a second-price auction, before which a bidder has the opportunity to  make a take-it-or-leave-it offer of a bribe in exchange for the opponent's absence from the auction (or bidding zero).    Since their work, alternative models in the same vein have been proposed. For example, \citet{Rach2013} considers the same bargaining protocol in first-price auctions and \citet{Rach2015} considers a model of alternating offers between two bidders with unequal bargaining power. \citet{Troyan2017}  extends ES's model to interdependent values and affiliated signals. 

The papers cited above share a common feature--namely, the collusion proposal specifies a positive transfer from the initiator to the opponent. However, it is natural that if offering a bribe to the opponent is feasible, then requesting a bribe from the opponent should also be available in the initiator's toolbox. 

Suppose
the collusion initiator commits to a double-option scheme--i.e., a
proposal consisting of both a bribe and a request. Compared with
the single-option scheme, such a double-option scheme seems to give a higher expected payoff for the initiator, provided it can be successfully
implemented. 

With the additional option, a multidimensional signaling game emerges and the problem of adverse selection may be more severe, so that  the existence and robustness of equilibria are at stake. 
To examine the implementability, we consider a second-price auction with two bidders, as in ES. Specifically, we  assume that a bidder has an opportunity to make a take-it-or-leave-it collusion proposal, consisting of both a bribe and a request, to    the opponent before   a second-price auction starts. The opponent can  accept at most one option in the proposal, in which case both bidders follow the proposal and bid cooperatively in the auction. If the opponent rejects the proposal, then both bidders compete noncooperatively in the auction.

The solution concept we use is  weak perfect Bayesian equilibrium (hereafter equilibrium). If an equilibrium survives   the D1 criterion 
(\citet{cho1990}),  the same standard refinement used in ES, it is said to be robust.\footnote{\citet{milgrom1986}
apply the intuitive criterion (\citet{cho1987}), which is weaker than the D1 criterion, for their multidimensional signaling model of  price and advertising expenditure decisions by a firm trying to convey information
about its product quality to consumers.} To illustrate our main points, we focus on  separating equilibria.  Although there may exist a multiplicity of separating equilibria, we show that there always exists a robust separating equilibrium in which the initiator's request
is equal to his valuation. In the equilibrium, the opponent always accepts the collusion proposal (accepting one of the two options), that is,  for each type of  initiator, the probability of a successful collusion is one. The significance of this result is threefold. First, a separating equilibrium does not exist in ES.\footnote{Log-concavity is assumed in \citet{ES2004}  to ensure the existence of nondecreasing equilibria. Common support for type distributions is assumed in \citet{ES2004}, which implies that an equilibrium, if it exists, is only nondecreasing, but not fully separating. If asymmetry in value support is allowed in their model, then a fully separating equilibrium may exist. }  Second,  in ES  the initiator typically cannot be guaranteed a successful collusion because the best response of the opponent is to accept the bribe if her type is low and reject  it otherwise. Hence, in our model such a double-option scheme leads to a qualitatively different prediction about the outcome of bargaining than that of  ES.  Third, and perhaps more importantly, the result shows that second-price auctions in this dynamic environment are as vulnerable as in the static model; namely, a successful collusion is guaranteed, even though now the bidders do not agree on collusion beforehand and both act strategically.\footnote{In \citet{Rach2015}, under some conditions, there exists an efficient equilibrium in which a successful collusion is also guaranteed. However, in that model   the two bidders engage in two rounds of alternating offers of bribes and thus the information leakage from the first round facilitates collusion. }  

Given  that the equilibrium always exists, it is interesting to investigate
the performance of the scheme from the perspective of the initiator. We next compare the expected payoff of the initiator in our model with the counterpart in ES. On one hand,  it seems that the additional option  alone represents an additional channel  through  which the initiator can  extract more surplus from the opponent. On the other hand, the additional option tends to intensify the adverse selection problem. An interesting feature of the separating equilibrium is that the bribe may exceed the initiator's valuation. In particular, the equilibrium bribing function in Proposition \ref{sep-equi} implies that if the lowest possible valuation of the initiator is zero, then the bribing function has an infinite derivative at zero. Hence, it may be very costly for some types of initiator to signal the strength. We show that in the identified robust equilibrium for each type
of   initiator, the expected payoff is generally higher than the
counterpart in any robust equilibria of  ES.

As mentioned above, in our model there may exist a multiplicity of separating equilibria. Hence, it is natural to ask whether there exist any other separating equilibria in which (at least) for some types of  initiator the expected  payoff can be even higher. It turns out that for each type of   initiator
 the expected payoff  in the identified equilibrium is also generally higher than the counterpart in any other separating equilibrium in our model. 

For convenience of exposition, our main results are derived in the absence of a reserve price in the auction. As explained in Section \ref{reserve}, a positive reserve price essentially does not change the results.

Although in this paper we focus on second-price auctions, we also discuss briefly the situation in which the auction format is changed to a first-price auction. It turns out that in this case, even though bidders have no incentive of on-path deviation, the initiator has an incentive to deviate off the path  and thus there are no separating equilibria if his lowest type is zero. This result is a reminiscence of  the ones in \citet{Rach2013} which considers a bribing model as ES but with a first-price auction. The driving forces for the nonexistence are also similar, i.e.,  the informational link between the bribing stage and the
bidding stage. On the other hand, in his model the equilibrium unravelling happens on the path, while in our model it happens off the path.  

The rest of the paper is organized as follows. This Section continues with
a brief review of related literature. In Section \ref{model} we describe the model. In Section \ref{results}  we characterize the separating equilibrium and show its robustness. In Section \ref{sect-payoff}  we compare the payoff of the initiator in our model with the counterpart in any robust equilibria in ES and  other separating equilibria in our model. In Section \ref{discussion} we  briefly discuss of the role of a reserve price and a switch to a first-price auction and Section \ref{conclusion} concludes. All proofs are relegated to the appendices.

\subsection*{Related literature}

Our paper contributes to the growing literature of collusion in auctions.
As stated in the introduction, the traditional approach is to consider
a static game before which some bidders have already agreed to participate
in
 collusion. On the other hand, ES-like dynamic models  typically
assume the single-option collusion proposal--i.e., an offer  of  a bribe from the bidder with the move to the opponent, in exchange for
the latter's cooperation. For example, apart from the papers cited in
the introduction, \citet{Chen2006} address the potential collusion problem of
the opponent  who cheats by using shill bidders in second-price auctions. \citet{Kivetz2010}
consider a complete information model of a first-price auction.  \citet{balzer2019}  considers a more general set of mechanisms for first price and second price
auctions. The main difference is that we consider
a simple and natural double-option proposal in a second-price auction and focus on its implementability and profitability
for the initiator, while Balzer's main objective is to examine the difference between a
first price and a second-price auction in terms of the possibility of efficient collusion. \citet{Zheng2019b} considers a model of first-price auctions before which a third party proposes a transfer between two bidders who then decide whether  to accept the proposed transfer. He characterizes the necessary and sufficient condition under which both bidders accept the proposed transfer almost surely (with respect to their priors).

More broadly, our paper is related to the strand  of mechanism-design literature that
studies the informed-principal problem, based on the seminal work of  \citet{myerson1983}; \citet{maskin1990}; and \citet{maskin1992}. In particular, \citet{maskin1990} consider a private value setting in which the principal proposes a contract
that itself is a game to be played when it is accepted by the agent, and 
the whole game ends if it is rejected. They show that the principal is generally
better off  compared with the scenario in which his type is of complete information.

 Our paper is also related to the literature of multidimensional signaling games pioneered by \citet{milgrom1986} and \citet{wilson1985}. In particular, \citet{milgrom1986} study the problem of a firm that can use both price and advertising expenditure
to signal its product quality to consumers.  In their model, the firm has  only two types  and the advertising expenditure is dissipative. In contrast, in our model the initiator
 has a continuum of types and neither of the  signals is dissipative.

\section{The model}\label{model} 

Two  risk-neutral bidders are about to attend a second-price auction in which there is a single indivisible object and no reserve price.\footnote{A  reserve price is discussed in Section \ref{reserve}.} Before the auction starts, bidder 1 (he) has an opportunity to make a take-it-or-leave-it proposal to bidder 2 (she). The proposal consists of a nonnegative bribe and a nonnegative request, denoted by $(b,r)$ with
$b,r\ge0$. If bidder 2 accepts the bribe, then  bidder 1 pays $b$ to her and she bids zero in the auction so that bidder 1 wins the object at a price of zero. If bidder 2 accepts  the request, then she pays $r$ to bidder 1, who then bids zero in the auction and she wins the object at a price of zero. If bidder 2 rejects the proposal, then both bidders bid non-cooperatively in the auction.

We assume bidders' valuation (or \textit{type}) distributions are independent, but allow for asymmetry. For $i=1,2$, bidder $i$'s valuation $v_i$ is independently distributed according to $F_i(v_i)$ on $[\underline v_i,\bar v_i]$.
  Each distribution function $F_i$ admits a continuous and differentiable density function $f_i(v_i)\in(0,\infty)$ for all $v_i$. We assume $\underline v_i  \ge 0$ and $\bar v_i <\infty$.

Our model is a dynamic game with incomplete information. We focus on pure strategy separating equilibria and thus both bidders do not randomize over actions  whenever they are supposed to take the move.  A pure strategy of bidder 1 specifies a proposal and a bid in case of competition in the auction for each type $v_1$; a pure strategy of bidder 2 specifies a choice from the choice set $ \mathbb{C}\equiv$ \{accept $b$, accept $r$, reject the proposal\} and a bid in case of competition in the auction for each type $v_2$. An equilibrium of the game is a profile of strategies accompanied by a belief system  of the bidders such that the strategies and the beliefs of both bidders  are rational and consistent with each other, both on and off the equilibrium path.\footnote{For a more formal definition of a weak perfect Bayesian equilibrium, see, e.g., \citet{gibbons1992}.} If an information set may be reached in the equilibrium, then it is said to be on the equilibrium path; if an information set can never be reached in the equilibrium, then it is said to be off the equilibrium path.\footnote{Because our game involves continuous type spaces, ``reached with positive probability" as in the standard definition for discrete type spaces is not appropriate here.} For each information set on the equilibrium path, bidders form their beliefs according to  Bayes's rule and the bidders' equilibrium strategies; for each information set off the equilibrium path, bidders form their beliefs according to Bayes's rule and the bidders' equilibrium strategies whenever possible. In particular, in an equilibrium of our game, the reason that bidder 1 has no incentives to deviate to an off-path proposal must be that following that proposal, there exist some beliefs of bidder 2 such that the resulting best responses of bidder 2 make it not profitable for any type of bidder 1 to deviate to that off-path proposal (compared with the equilibrium payoffs). On the other hand, upon receiving an off-path proposal, bidder 2 can form arbitrary beliefs because Bayes's rule does not apply there. But  once an off-path belief is formed (which should then be the common belief for both bidders),    the optimal choices from $\mathbb{C}$ and the optimal bid in case of competition in the auction must be consistent with the belief  and bidder 1's  best responses in the auction.

We assume that   if bidder 2 rejects the proposal, then both bidders   bid truthfully in the second-price auction since it is a weakly dominant strategy in the auction.

We let $b(v_1)$ and $ r(v_1)$ be the bribing and requesting functions in an equilibrium.
An immediate observation of our model is that if $\underline v_1\ge\bar v_2 $,   there always exists a separating equilibrium in which $b(v_1)=0$ and $r(v_1)=v_1$. In the equilibrium, all types of bidder 2 accepts the zero bribe and bidder 1 realizes his valuation.\footnote{The single-option model in ES shares the same equilibrium in this case, i.e., $b(v_1)=0$ for all $v$.  } Hence, below we focus on the nontrivial case of $\underline v_1<\bar v_2$.

\section{A separating equilibrium}\label{results}

Suppose there exists a separating equilibrium in which  bidder 1's request is not greater than his own valuation. We assume that bidder 2 is willing to accept a proposal if  she is indifferent between accepting and rejecting it.
In the equilibrium, upon receiving a separating proposal $(b,r)$ from type $v_1$, for bidder 2, accepting $b$ gives a payoff of $b $, accepting $r$ gives a payoff of $v_2-r$,  and rejecting the proposal gives a payoff of $v_2-v_1$. Thus, if $r\le v_1$, it is optimal for
bidder 2 to accept $b$ if $v_2\le b+r$ and accept $r$ if  $v_2> b+r$.\footnote{Although type $v_2=b+r$ is indifferent between accepting $b$ and accepting $r$, for simplicity we let her accept $b$ when she accepts the proposal.} That is, if
bidder 1 requests no more than his own valuation, then the best response
of bidder 2 is to always agree to collude. If her valuation is low, she accepts the bribe; if her valuation is high, she accepts the request. 
These characterizations lay the foundation for our subsequent analysis of the collusive equilibrium. Following the cutoff strategy
of bidder 2,   
the expected payoff of bidder 1 with valuation $v_1$ from a separating proposal
$(b,r)$ with $r\le v_1$  is  $\pi(v_1,b,r)=F_2(b+r)(v_1-b)+ (1-F_2(b+r))r$, which can be rewritten as
\begin{equation*}
\pi(v_1,b,r)= F_2(b+r)(v_1-(b+r))+r.
\end{equation*}
 
For convenience of exposition, let $v^1$ be the smallest value such that, if it exists,
\begin{equation}
b(v^1)+v^1=\bar v_2,
\end{equation}
and thus $F_2(b(v^1)+v^1)=1$. To save notation,  we let $\pi(v_1)\equiv\pi(v_1,b(v_1),r(v_1))$.

For any type $v_1< v^1$, the on-path  incentive compatibility (IC) condition for an equilibrium in which $r(v_1)\le v_1$ is  
\begin{align}\label{ic-general}
v_1\in \arg\max_{t}  \; \pi(v_1,b(t),r(t)) = F_2(b(t)+r(t))(v_1-(b(t)+r(t)))+r(t),
\end{align}
which implies
\begin{equation} \label{ic-general1}
[f_2(b(v_1)+r(v_1))(v_1-(b(v_1)+r(v_1)))-F_2(b(v_1)+r(v_1))](b'(v_1)+r'(v_1))+r'(v_1)=0.
\end{equation}

The following proposition shows that for any distribution $F_1$ and $F_2$, there always exists a separating equilibrium in which  bidder 1 requests his own valuation.

\begin{proposition}\label{sep-equi} 
There exists a separating equilibrium in which
\begin{itemize}
\item
 for any type  $v_1\le v^1 $,  $b(v_1)$ is given by 
\begin{align}\label{ic-r-eq-v}
b'(v_1)=\frac{1 }{f_2(b(v_1)+ v_1)b(v_1) +F_2(b(v_1)+v_1)} -1,
\end{align}
with an initial condition $b(\underline v_1)=\max\{\underline v_2-\underline v _1,0 \}$, which always admits a unique solution.  For any type $v_1>v^1$, $b(v_1)=b(v^1)$.\footnote{Because \eqref{ic-r-eq-v} implies $b(v_1)+v_1$ is strictly increasing,  $b'(v_1)=0$
for all $v_1>v^1$. }

\item 
for any type $v_1$, $r(v_1)=v_1$.\footnote{Technically, for any type $v_1>v^1$, $r(v_1)$ can be any value such that $b(v^1)+r(v_1)\ge \bar v_2$ and $r(v_1)$ takes different values for different $v_1$. Alternatively, in an equilibrium, for any type $v_1>v^1$, $r(v_1)$ can be equal to $  v^1$ so that there is a pooling segment on the top. However, the equilibria are essentially equivalent.  } 

\item
upon receiving any given  equilibrium proposal $(b(v_1),v_1)$, bidder 2 always accepts it, i.e., accepts $b(v_1)$ if $v_2\le b(v_1)+v_1$ and accepts the request $v_1$ if $v_2>b(v_1)+v_1$. 

\end{itemize}
 
\end{proposition}
\begin{proof}
See Appendix \ref{prof-sep-equi}.
\end{proof}

Below, we show that there exists an equilibrium with the same on-path behavior as in Proposition \ref{sep-equi}  and a system of reasonable off-path beliefs such that given the resulting best
responses of bidder 2, it is not profitable for bidder 1 to deviate to any  off-path proposals. In particular, the off-path beliefs of bidder 2 are reasonable in the sense of the D1 criterion.

Cho and Sobel's D1 criterion says that if, upon observing  an  off-path action,  any  best responses of the receiver (based on some beliefs about the sender's type) that imply a  profitable  deviation for a sender type also imply  a profitable deviation for another sender type and the converse is not true, then, upon observing that off-path action,  the receiver's posterior beliefs should place zero probability on the former type. Roughly speaking in an alternative way, if a type of the sender is dominated by another type in terms of the expected payoff of the sender for possible best responses of the receiver, then the former type should be excluded from the set of reasonable beliefs of the receiver. Hence, if for an off-path action there exists some type of the sender that is not dominated by any other type, then a reasonable belief of the receiver is that it is the sender's type.   If for each off-path action there exists such a reasonable belief that the best response of the receiver based on the belief implies that it is not profitable for any type of  sender to deviate to the off-path action, then the equilibrium survives the D1 criterion.

We first describe the set of best responses of bidder 2 upon receiving an off-path proposal $(b,r)$. 
\begin{lemma}\label{best-resp-2}
Upon receiving  an  off-path  proposal $(b,r)$, the best response of bidder 2 can be described as a pair of critical values $(v_2^b,v_2^r)$ such that she accepts $b$ if $v_2\le v_2^b$, rejects the proposal if $v_2^b<v_2< v_2^r$, and accepts $r$ if $v_2\ge v_2^r$. In particular, 
\begin{equation}
 v_2^b \le \min\{b+r,\bar v_2\} \le v_2^r\le \bar v_2.
\end{equation}
  
\end{lemma}
\begin{proof}
See Appendix \ref{proof-best-resp-2}.
\end{proof}
 
For an off-path proposal $(b,r)$, given a best response $(v_2^b,v_2^r)$  of bidder 2 as described in Lemma \ref{best-resp-2}, the expected payoff of type $v_1$, denoted by $\pi(v_1,b,r,v_2^b,v_2^r)$,   is given by
\begin{align}\label{payoff-off}
\hspace{-1em}\pi(v_1,b,r,v_2^b,v_2^r)&\equiv F_2(v_2^b)(v_1-b)+\int_{\min\{v_2^b,v_1\}}^{\min\{v_2^r,v_1\}} (v_1-v_2)f_2(v_2) dv_2+(1-F_2(v_2^r))r.
\end{align}
For convenience, we record the following fact that follows directly from the envelope theorem and \eqref{ic-r-eq-v}.\footnote{The proofs of the facts
are given in Appendix \ref{prof-facts}.}

\begin{fact}\label{convexPi}
$\pi'(v_1)=F_2(b(v_1)+v_1)$, $\pi(v_1)=v_1-F_2(b(v_1)+v_1)b(v_1)$ and $b(v_1)+v_1$ is strictly increasing. Type $\underline v_1$ earns his valuation, i.e., $\pi(\underline v_1)=\underline v_1$.
\end{fact}


\begin{fact} \label{aboveV2b}
For an off-path proposal $(b,r)$ and a best response $(v_2^b,v_2^r)$ of bidder 2,   if $\pi(v_2^b,b,r,v_2^b,v_2^r)\le \pi(v_2^b)$, then for any type $v_1> v_2^b$, 
$\pi(v_1,b,r,v_2^b,v_2^r)< \pi(v_1).$
\end{fact}

The next two  facts identify some off-path proposals that would
never be made by any type of bidder 1. 
\begin{fact} \label{rAboveV}
Consider an off-path proposal $(b,r)$. If $r\le \underline v_1$, then it is not profitable for any type $v_1$ to deviate to $(b,r)$ for any belief of bidder 2. 
\end{fact}

Intuitively, since bidder 1's value for bidder 2 is at least   $\underline v_1$,   bidder 1 does not have an incentive to request an amount lower than $\underline v_1$.  Fact \ref{rAboveV} implies that we can restrict  attention to proposals with $r>\underline v_1$. 

The following fact further excludes  proposals $(b,r)$  with $r>\underline v_1$ and $b\ge \bar v_2- \underline v_1$.

\begin{fact} \label{bAboveV}
Consider an off-path proposal $(b,r)$. If $r>\underline v_1$ and $b\ge \bar v_2- \underline v_1$, then it is not profitable for any type $v_1$ to deviate to $(b,r)$ for any belief of bidder 2. 
\end{fact}

Fact \ref{rAboveV}  and Fact \ref{bAboveV} together imply that we can restrict  attention to proposals with $r>\underline v_1$ and $b< \bar v_2- \underline v_1$.

\begin{fact}\label{piBelow}
Consider an off-path proposal $(b,r)$ with $r>\underline v_1$ and $b<\bar v_2- \underline v_1$. If $b>0$ or $\underline v_1>0$,  then 
$\pi(v_1,b,r,b+\underline v_1,\bar v_2)< \pi(v_1)$ 
for any type $v_1$.
\end{fact}

We are now ready to show that there exists an  equilibrium with the same on-path behavior as in Proposition \ref{sep-equi}  and it survives the D1 criterion.

\begin{proposition}\label{robust}
There exists an equilibrium that shares the same bribing function and requesting function as in Proposition \ref{sep-equi} and survives the D1 criterion.
\end{proposition}

\begin{proof}
See Appendix \ref{robust-prof}.
\end{proof}

The following example sheds  light on the equilibrium.

\begin{figure} 
   \begin{center} 
  \includegraphics[scale=0.445]{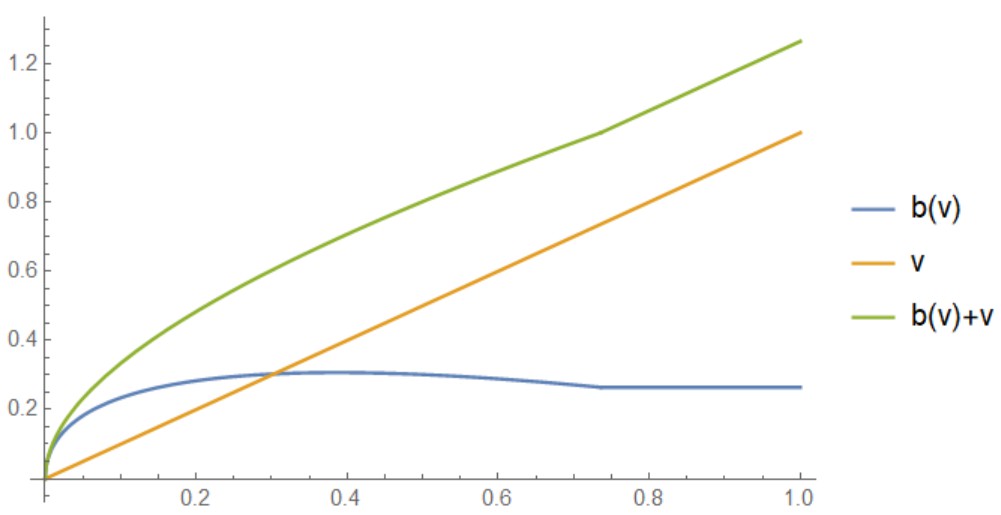}
   \end{center}
   \caption{The equilibrium bribing and requesting functions when $F_2(x)=x$ and $\underline v_1=0$.}
   \label{solution-b(v_1)} 
   \end{figure}

\begin{example}
Suppose $F_2(x)=x$ on $[0,1]$.  Suppose $v_1$ is distributed on $[0,1]$. Then for all $v_1$ satisfying $b(v_1)+v_1\le 1$,  \eqref{ic-r-eq-v} becomes 
\begin{equation*}
b'(v_1) = \frac{1}{2b(v_1)+v_1}-1.  
\end{equation*}
The solution is 
\begin{equation*}
b(v_1)=\frac{1}{2} \left(2 W\left(-e^{-\frac{v_1}{2}-1}\right)-v_1+2\right),
\end{equation*}
where $W(x)$ solves
\begin{equation*}
x=W(x)e^{W(x)}.
\end{equation*}

With the initial condition $b(0)=0$,  the equilibrium
bribing function $b(v_1)$  is plotted  in Figure \ref{solution-b(v_1)}. \end{example}  

The example above illustrates two interesting features of the
equilibrium bribing function, which help to distinguish our model from that of ES. 
 
First, unlike in the single-option model of ES, here, for some
low types of bidder 1, the bribe may exceed his valuation in the equilibrium. Intuitively,
in the example, for low types of bidder 1, both the bribes and the requests are low (in the sense of absolute magnitude).
Upon receiving a proposal of low bribe and low request, with high probability bidder 2 would decline
the bribe and accept the request.
Since the bribe is accepted with a low probability, for bidder 1 it
is a ``low cost'' signal. The low types of bidder 1 are thus  incentivized to gamble with a bribe that is potentially higher than his valuation to signal their values.
On the other hand, type zero of bidder 1 earns zero in the equilibrium and his expected payoff from mimicking
some other type $v_1>0$ is $\pi(0,v_1)=-(b(v_1)+v_1)b(v_1)+(1-(b(v_1)+v_1))v_1$, which is equivalent to $v_1-(b(v_1)+v_1)^2 $.
Thus, in order for $\pi(0,v_1) $ to be nonpositive, $b(v_1)$ should be at least as large as $\sqrt {v_1} -v_1$, which exceeds $v_1$
for small $v_1$. That is, the incentive compatibility of type zero actually requires that   the bribes  of low
types of bidder 1 exceed their values.
 
Second, the bribing function can be nonmonotonic for high types of bidder 1, while the bribing function in ES must be nondecreasing. In our model, both the
bribe and the request are costly signals. As stated above, for
low types of bidder 1, their requests are low and thus will be accepted with high probabilities.
These low types of bidder 1 need to rely more on relatively high but less costly bribes for signalling.
For higher types of bidder 1, the probabilities of their requests being accepted are low,  and
thus the request becomes a less costly signal. With a higher request being offered together, a bribe is accepted with a high chance and thus becomes a relatively more costly signal.
As the value of bidder 1 gets higher, his request gets higher and his bribe gets more costly. The higher types of bidder 1  can thus  have less incentive to offer higher bribes, which makes it possible that the bribing function can even decrease in types. Indeed, the differential equation in \eqref{ic-r-eq-v} implies that for type $v_1=v^1$ (if it exists), we must have $b'(v^1)=\frac{1}{f_2(\bar v_2)b(v^1)+1}-1<0$. In other words, we must have a locally decreasing bribing function to the left of $v_1=v^1$.

\section{The expected payoff of the initiator}\label{sect-payoff}

In this section we first compare the expected payoff of bidder 1 in the identified robust equilibrium in our model with the counterpart in ES. 

In ES, bidder 1 commits to offering a variable take-it-or-leave-it bribe to bidder 2. For separating bribes, let  the bribing function be $B_{es}(v_1) $ in their model.  Then the incentive compatibility condition for type $v_1$ with a separating  bribe is
\begin{align*}
v_1\in\arg\max_t\ \Pi(v_1,t)=F_2(B_{es}(t)+t)(v_1-B_{es}(t))+\int_{\min\{B_{es}(t)+t,v_1\}}^{v_1} (v_1-x) f_2(x)dx,
\end{align*}
which implies 
\begin{equation}\label{es-cont-b}
B_{es}'(v_1)= \frac{f_2(v_1+B_{es} (v_1))(v_1-B_{es} (v_1))}{F_2(v_1+B_{es} (v_1))-f_2(v_1+B_{es} (v_1))(v_1-B_{es} (v_1))}.
\end{equation}
Under some regularity conditions, ES identify the set of  equilibria that survive the D1 criterion. In such an equilibrium, for some $\hat v_1$, the bribing function, denoted by $B(\cdot)$, is 
\begin{equation}
B(v_1) =
\begin{cases}
B_{es}(v_1)  &\text{ if \;}  v_1<\hat v, \\
\hat B\equiv\hat v_1-F_2(\hat v_1+B_{es}(\hat v_1))(\hat v_1-B_{es}(\hat v_1))  &\text{ otherwise, } 
\end{cases} \label{Bhat}
\end{equation}
where  $\hat B\ge \bar v_2-E[v_1|v_1\ge \hat v]$. 
The expected payoff function of bidder 1, denoted by $\Pi(\cdot)$, in such an equilibrium is  
\begin{equation}\label{es-payoff}
\Pi(v_1) =
\begin{cases}
F_2(v_1+B (v_1))(v_1-B (v_1))  &\text{ if }\ v_1\le \hat v, \\
v_1-\hat B  &\text{ otherwise. }
\end{cases}
\end{equation}

\begin{proposition}\label{sep-pi-vs-es}
Suppose there exists a robust equilibrium in ES.  Then $\pi(v_1)\ge\Pi(v_1)$ for any $v_1$.
\end{proposition}

\begin{proof}
See Appendix \ref{prof-sep-pi-vs-es}.
\end{proof}

So far, we have focused on the separating equilibrium in which bidder 1's request is his valuation. Clearly, there may be some other pairs of $(b(v_1),r(v_1))$, with $r(v_1)\le v_1$, that satisfy the incentive compatibility condition in \eqref{ic-general1} and thus may sustain a separating equilibrium if  $(b(v_1),r(v_1))$ is separating. Moreover, there may also exist some separating equilibria in which,  for some type $v_1$, $r(v_1)>v_1$ so that the request is never accepted. A natural question is whether the expected payoff of bidder 1 can be improved in some of those separating equilibria against the equilibrium identified above. We show below that for each type of bidder 1, the expected payoff in  any separating equilibrium in which $r(v_1)\ne v_1$ is generally lower than the counterpart in the   equilibrium identified above. 

\begin{proposition}\label{other-sep}
Suppose that in our model there exists a different separating equilibrium with  a pair of bribing and requesting functions $(\beta(v_1),\gamma(v_1))$. Let the  expected payoff of bidder 1 in the different equilibrium be $\pi(v_1;\beta,\gamma)$. Then $\pi(v_1)\ge \pi(v_1;\beta,\gamma)$.
\end{proposition}
\begin{proof}
See Appendix \ref{prof-other-sep}.
\end{proof}

\section{Discussion}\label{discussion}

\subsection{Reserve price}\label{reserve}

We note that the previous results remain essentially unchanged even if there is a reserve price in the auction. Specifically, given a positive reserve price, there exists a robust separating equilibrium in which, conditional on the object being sold--i.e., at least one of the bidders' valuations exceeds the reserve price--the   collusion success probability is one. The intuition is the following. Let the reserve price be $\mathcal{R}$. Observe that any type of bidder 1 below $\mathcal{R}$ is not interested in winning the object and he has no value for bidder 2. In the separating equilibrium similar to the above, these types are correctly identified as below $\mathcal{R}$. So for any  type $v_1< \mathcal{R} $, the equilibrium bribe $b(v_1)=0$, which is never accepted by any $v_2>  \mathcal{R}$, and any  request  $r(v_1)>0$ ($=0)$ will be rejected (accepted) by any type of bidder 2. On the other hand, for any type $v_1\ge\mathcal{R}$, the equilibrium request  is $r(v_1)=v_1-\mathcal{R}$, i.e., the valuable part for bidder 2, since if bidder 2 accepts the request, she still needs to pay $\mathcal{R}$ to the auctioneer.

More formally, when a proposal $(b,r)$ from type $v_1$ is separating and $r\le v_1-\mathcal{R}$, then bidder 2 accepts $b$ if  $v_2\le b+r+\mathcal{R}$ and accepts $r$ if  $v_2> b+r+\mathcal{R}$. So the expected payoff of bidder 1 with type  $v_1$ is $\pi(v_1)=F_2(b+r+\mathcal{R})(v_1-b-\mathcal{R})+(1-F(b+r+\mathcal{R}))r$, which can be rewritten as
\begin{equation*}
\pi(v_1)=F_2(b+r+\mathcal{R})(v_1-(b+r+\mathcal{R}))+r.
\end{equation*}

The IC condition
for the separating equilibrium with $r(v_1)\le v_1-\mathcal{R}$  becomes
\begin{align*}
&[f_2(b(v_1)+r(v_1)+\mathcal{R})(v_1-(b(v_1)+r(v_1)+\mathcal{R}))
 \\
&\quad -F_2(b(v_1)+r(v_1)+\mathcal{R})](b'(v_1)+r'(v_1))+r'(v_1)=0.
\end{align*}
In the equilibrium with  $r(v_1)=v_1-\mathcal{R}$ for type $v_1\ge \mathcal{R} $,  it is the same as \eqref{ic-r-eq-v}. The only difference is that the initial condition is changed into 
$b(\max\{\underline{v}_{1},\mathcal{R}\})=\max\{\underline{v}_{2}-\max\{\underline{v}_{1},\mathcal{R}\},0\}$
and $r(\max\{\underline{v}_{1},\mathcal{R}\})=\max\{\underline{v}_{1},\mathcal{R}\}-\mathcal{R}$.\footnote{So $b(\max\{\underline{v}_{1},\mathcal{R}\})+r(\max\{\underline{v}_{1},\mathcal{R}\})+\mathcal{R}\ge \underline v_2$. Again, since $b(v_1)+r(v_1)+\mathcal{R}$ is strictly increasing in $v_1$, by  similar arguments in the proof of Proposition 1, the IC condition admits a unique solution for all $v_1\ge \max\{\underline{v}_{1},\mathcal{R}\} $. For any $v_1<\max\{\underline{v}_{1},\mathcal{R}\}$, we require $b(v_1)=r(v_1)=0$. Observe that only if $\underline v_1<\mathcal{R}$, will there be a positive measure of type $v_1<\mathcal{R}$. In that case, $r(\mathcal{R})=0$ and for any $v_1>\mathcal{R}$, $r(v_1)>0$. Because any type $v_1<\mathcal{R}$ is not interested in winning the object and he has no value for bidder 2, he earns zero payoff in the equilibrium. Any type $v_1<\mathcal{R}$ has no incentive to mimic any type $v_1>\mathcal{R}$ because by the IC condition, type $v_1=\mathcal{R}$ has no incentive to do so.     } 
 Hence, the equilibrium proposal of any type of bidder 1 is always accepted by bidder 2 (although a zero request has no value for her  if her valuation is not higher than the reserve price).

\subsection{First-price auction}\label{first-price}

In this section we briefly discuss the situation in which the auction format is changed to a first-price auction.

  \citet{Rach2013} considers the same single-option scheme as in \citet{ES2004}, but  with first-price auctions. The important result from his model is that the existence of a separating equilibrium is generally impossible and there may even be no pooling equilibrium. In his model, an important feature of the  (pure strategy) equilibrium of the continuation games is that when  a proposal reveals the initiator's type perfectly and is rejected,  the initiator bids his valuation, $v_1$, and the rejectors submit  the ``minimally winning" bid if they find winning worthwhile--i.e., a value of $v_1^+$, which wins with certainty against any $v_1'\le v_1$  (but pays $v_1$) and loses against any
$v_1'>v_1$. The driving force for his results of the nonexistence of equilibrium is that a high-type initiator has the incentive to cheat the opponent by mimicking a low type $v_1$, because on the path the bribe $b(v_1)$ of the low type $v_1$ can be rejected, and once it is rejected, the high type can bid a value marginally higher than $v_1^+$ and win the auction at a low price. 

Suppose now that a hypothetical
separating equilibrium exists in the double-option model for a first price
auction. In that case, we note that if, in the hypothetical equilibrium, bidder 1
believes that the rejector types of his separating proposal are all higher
than his type, then truthful bidding after being rejected is a best response (in the same sense
as in \citet{Rach2013}). This then implies that upon receiving a separating
proposal, the best response of the opponent would be the same as the one in
the case of a second-price auction. Hence, the expected payoff of the
initiator would also be the same as the one in the case of a second-price auction, which means that the on-path incentive compatibility condition is
satisfied as in the case of a second-price auction. Since in the hypothetical
separating equilibrium the separating proposals would always be accepted on
the path, the continuation games would never be played and thus become
off-path events. Hence the driving force for the nonexistence of a
separating equilibrium in \citet{Rach2013} would disappear on the path in
the hypothetical equilibrium in the double-option model.

However, as shown below, when the lowest type of bidder 1 is zero, i.e.,
bidder 1 may have no value for the object, no separating equilibria with full acceptance
exist.

\begin{proposition}\label{nonexistence-1st}
Suppose the auction format is first-price auction. Suppose bidder 1's valuation is distributed on $[0,\bar v _1]$ and bidder 2's valuation is independently distributed on $[0,\bar v_2] $. There exist no separating equilibria in which bidder 1's proposals are always accepted. 
\end{proposition}

\begin{proof}
See Appendix \ref{prof-nonexistence-1st}. 
\end{proof}

 The reason for such a nonexistence result for the case of a first-price auction is similar to that of \citet{Rach2013}. In a first-price auction, the equilibrium unravelling is due to the strategic link between the bargaining stage and the bidding stage. In particular here, if the opponent rejects a proposal, then she will not bid more than the requested amount at the bidding stage (otherwise it would be better to accept the request at the bargaining stage). In a second-price auction, it remains a weakly dominant strategy for the opponent to bid truthfully, and thus the initiator is not able to trick the opponent at the bargaining stage. As shown in the proof of Proposition \ref{nonexistence-1st}, there is an obvious difference here; in \citet{Rach2013} unravelling happens on the path, while in our model it happens off the path. A more complete analysis of
the first-price auction scenario is beyond the scope of
this paper, and we leave that for future work.

\section{Conclusion}\label{conclusion} 

We  examine a collusion model for second-price auctions in which a bidder has the opportunity to propose a combination of an offer and a request of bribes to the other bidder. The bidders are involved in a multidimensional signaling game. Even when asymmetry is allowed, we show that the collusion initiator with full bargaining power can always secure a successful collusion in such a dynamic environment, as in previous literature with a static environment.  This result confirms the susceptibility of second-price auctions to collusion in dynamic environment, but has a qualitatively different nature than the previous dynamic model with a single-option considered by \citet{ES2004}. Finally, we show that for
each type of  initiator the payoff from such a double-option scheme
is generally higher than the counterpart from the single-option scheme.

\appendix
\section*{Appendix}

\section{Proof of  Proposition \ref{sep-equi} }\label{prof-sep-equi}

In the equilibrium with $r(v_1)=v_1$, the bribing function $b(v_1)$ satisfies
\begin{equation*}
[ f_2(b(v_1)+ v_1)b(v_1) +F_2(b(v_1)+v_1)](b'(v_1)+1)=1, 
\end{equation*}
which in turn can be rewritten as \eqref{ic-r-eq-v}.

When $\underline{v}_{1}<\underline{v}_{2}$, $b(\underline{v}_{1})=\max\{\underline{v}_{2}-\underline{v}_{1},0\}=\underline{v}_{2}-\underline{v}_{1}>0$
and  $b'(\underline{v}_{1})=1/[f_{2}(\underline{v}_{2})(\underline{v}_{2}-\underline{v}_{1})]-1$, which is finite.
Observe that although  $b'(\underline{v}_{1})$ can be negative and
thus $b(v_{1})$ is decreasing  in the neighborhood of $\underline{v}_{1}$,
$b(v_{1})$ is strictly positive in the neighborhood of $\underline{v}_{1}$. Furthermore, \eqref{ic-r-eq-v} implies $b(v_1)+v_1$ is strictly increasing for all  $v_1$. Thus, for all $v_1>\underline v_1$, $b(v_1)+v_1>\underline v_2$ and thus $F_2(b(v_1)+v_1)>0$. So for all $v_1>\underline  v_1$, the denominator in \eqref{ic-r-eq-v}  is strictly positive and thus the fraction is finite. Thus   $b'(v_1)$ is finite.  On the other hand, whenever $b(v_{1})$ is small
enough (but positive) such that the denominator in \eqref{ic-r-eq-v}  is smaller
than one for some $v_{1}>\underline{v}_{1}$, $b'(v_{1})$ is positive and
finite for $v_1<v^1$ and thus $b(v_1)$ bounces back to higher positive values.  So $b(v_1)>0$ for all $v_1$. Together we can now conclude that in this case \eqref{ic-r-eq-v} admits a unique solution.\footnote{That is, the usual arguments for the existence and uniqueness of a solution--e.g., Theorem 24.5 in 
\citet{simon1994}--apply.} 

If $\underline v_1>\underline v_2$, then by the similar arguments above (the bouncing-back property of $b(v_1)$ and finiteness of $b'(v_1)$), it
is straightforward to see that \eqref{ic-r-eq-v} also admits a unique solution. 

Suppose now that  $\underline{v}_{1}=\underline{v}_{2}$.  In this case
we have the initial condition $b(\underline{v}_{1})=0$
and thus $b'(\underline{v}_{1})=\infty$. This implies in the neighborhood of $\underline v_1$, $b(v_1)$ is strictly increasing; thus for each candidate local solution there exists an inverse function, denoted by $v_1(b)$, with zero derivative which is finite. Consider the inverse function version of \eqref{ic-r-eq-v}, given by\footnote{
 $v_1(b)$ is  the inverse function of $b(v_1)$.}
\begin{equation}
v_1'(b)=\frac{f_2(b+v_1(b))b+F_2(b+v_1(b))}{1-[f_2(b+v_1(b))b+F_2(b+v_1(b))]},
\end{equation}
 which  admits a unique local solution with the initial condition $v_1(0)=\underline v_1$, in the neighborhood of zero.  Hence,  \eqref{ic-r-eq-v}
admits a unique local solution in the neighborhood of $\underline v_1$. Then,  again
by the similar arguments above (the bouncing-back property of $b(v_1)$ and finiteness of $b'(v_1)$), the right side of \eqref{ic-r-eq-v} is well defined
and continuous
for all $b(\cdot)>0$ and $v_1>\underline v_1$. So \eqref{ic-r-eq-v} also admits a unique solution throughout the support of $F_1$.\footnote{ Again, \eqref{ic-r-eq-v} implies $b(v_1)+v_1$ is strictly increasing for all  $v_1$.  So $f_2(b(v_1)+v_1)=0$ and $F_2(b(v_1)+v_1)=1$ for all $v_1 > v^1$. It follows from \eqref{ic-r-eq-v}  that for any type $v_1>v^1$, $b'(v_1)=0$ and  $b(v_1)=b(v^1)$. Intuitively,  for these types the equilibrium proposals must be such that    $b(v_1)$  is always accepted by all types of bidder 2 and the equilibrium payoff of such types is $v_1-b(v^1)$.  }

Because of the positiveness of $b(v_1)$, the solution is an admissible bribing function.
The equilibrium is separating because the requesting function $r(v_1)=v_1$ is strictly increasing. 

Observe that in any case above, type $\underline v_1$'s request is accepted with probability one and thus he earns his valuation $\underline v_1$. For any type $v_1>\underline v_1$, $b(v_1)+v_1>\underline v_2$, so the bribe is accepted with positive probability and is a costly signal. Then the envelope theorem implies $\pi'(v_1)=F_2(b(v_1)+v_1)$, and thus $\pi(v_1)\ge0$ for all $v_1$.     

Thus, the bribing function and the requesting function described in the proposition  also satisfy the interim individual rationality condition. Next, we show that they satisfy on-path incentive compatibility.

We now show that there is no profitable on-path deviation for any type $v_1$--i.e., given that all other types of bidder 1 are following the equilibrium proposals, it is not profitable for type $v_1$ to send any on-path proposal $(b(t),t)$, $t\neq v_1$. Consider  the expected payoff of type $v_1$ from deviating to   $(b(t),t)$, i.e., $\pi(v_1,b(t),t)$. While 
$\pi_{v_1}'(v_1,b(t),t)=F_{2}(b(t)+t)$ for any given $t$, the envelope theorem implies that $\pi'(v_1)=F_{2}(b(v_1)+v_1)$. Because $b(v_1)+v_1$ is strictly increasing, this implies that for any given  $t$, $ \pi'(v_1)-\pi_{v_1}'(v_1,b(t),t)\gtrless 0$ for any $v_1\gtrless t$. Hence $\pi(v_1)>\pi(v_1,b(t),t)$ for any $v_1\neq t$. 

In the equilibrium, as explained above, upon receiving an on-path proposal, clearly it is not profitable for any type $v_2$ to deviate on the path--i.e., it is not profitable for type $v_2 \le b(v_1)+v_1$ to accept $r$ nor for type $v_2 >b(v_1)+v_1$ to accept $b(v_1)$. Thus the continuation games are never played and become off-path events. We next turn to off-path deviations. 

There are two types of off-path deviations in the equilibrium. First, there are many unsent off-path proposals by bidder 1. Second, because in the equilibrium, given an on-path proposal, bidder 2 on the path is supposed to accept the proposal with probability one, rejection becomes an off-path deviation as well. However, for the latter type of off-path deviations, it is straightforward to see that for any belief of bidder 1 it is not profitable for any type of bidder 2 to reject the received equilibrium proposal, since it is a weakly dominant strategy for bidder 1 to submit a bid of his valuation in the auction. So the only nontrivial type of off-path deviations is the former.

A simple belief to prevent the off-path deviation
of bidder 1 is the lowest type belief, i.e., for any off-path proposals,
bidder 2 believes that bidder 1's type is $\underline{v}_{1}$. To
see this, consider an arbitrary off-path proposal $(b,r)$. 

If the request is higher than $\underline{v}_{1}$, then with a belief
$v_{1}=\underline{v}_{1}$ the best response of bidder 2 is to accept
the bribe if $v_{2}\le  b+\underline{v}_{1}$ and reject the proposal
otherwise and then bid truthfully in the auction. The expected payoff
of type $v_{1}$ is then 
\[
\pi(v_{1},b,r;\underline{v}_{1})=F_{2}(  b+\underline{v}_{1})(v_{1}-b)+\int_{\min\{  b+\underline{v}_{1},v_{1}\}}^{\min\{\bar{v}_{2},v_{1}\}}(v_{1}-v_{2})f_{2}(v_{2})dv_{2}.
\]
Thus 
\[
\pi_{v_{1}}'(v_{1},b,r;\underline{v}_{1})=\begin{cases}
F_{2}(  b+\underline{v}_{1}) & \;\mathrm{if}\;v_{1}\le  b+\underline{v}_{1},\\
F_{2}(v_{1}) & \;\mathrm{if}\;v_{1}>  b+\underline{v}_{1}.
\end{cases}
\]
 Therefore, $\pi(v_{1},b,r;\underline{v}_{1})$ is linear for $v_{1}\in[\underline{v}_{1},  b+\underline{v}_{1}]$
and convex on $v_{1}\ge  b+\underline{v}_{1}$. Recall that $\pi(v_{1})$
is convex and $\pi'(v_{1})=F_{2}(b(v_{1})+v_{1})$.
For $v_{1}\in[\underline{v}_{1},  b+\underline{v}_{1}]$, $\pi(v_{1})-\pi(v_{1},b,r;\underline{v}_{1})$
is minimized at some $\hat{v}_{1}$ at which $\pi_{v_{1}}'(\hat{v}_{1},b,r;\underline{v}_{1})
=F_{2}(  b+\underline{v}_{1})=F_{2}(b(\hat{v}_{1})+\hat{v}_{1})=\pi'(\hat{v}_{1})$.
Since $\hat{v}_{1}\ge\underline{v}_{1}$, we have $b(\hat{v}_{1})\le b$,
which implies $\pi(\hat{v}_{1},b,r;\underline{v}_{1})=F_{2}(  b+\underline{v}_{1})(\hat{v}_{1}-b)\le F_{2}(b(\hat{v}_{1})+\hat{v}_{1})(\hat{v}_{1}-b(\hat{v}_{1}))+(1-F_{2}(b(\hat{v}_{1})+\hat{v}_{1})))\hat{v}_{1}=\pi(\hat v_1)$.
Consequently, for all $v_{1}\in[\underline{v}_{1},  b+\underline{v}_{1}]$,
$\pi(v_{1},b,r;\underline{v}_{1})\le\pi(v_{1})$. On the other hand,
since for $v_{1}>  b+\underline{v}_{1}$, $\pi_{v_{1}}'(v_{1},b,r;\underline{v}_{1})=F_{2}(v_{1})\le F_{2}(b(v_{1})+v_{1})=\pi'(v_{1})$,
we have $\pi(v_{1},b,r;\underline{v}_{1})\le\pi(v_{1})$ as well.

If the request is below or equal to $\underline{v}_{1}$,
then with the  belief $v_{1}=\underline{v}_{1}$ the
best response  of bidder 2 is to accept $b$ if $v_2
\le b+r$  and accept $r$ otherwise. The
expected payoff of bidder 1 is thus $\pi(v_1,b,r;\underline v_1)=F_2(b+r)(v_1-b)+(1-F_2(b+r))r$, which is linear in $v_1$. Again, $\pi(v_{1})-\pi(v_{1},b,r;\underline{v}_{1})$
is minimized at some $\hat{v}_{1}$ at which $\pi_{v_{1}}'(\hat{v}_{1},b,r;\underline{v}_{1})
=F_{2}(b+r)=F_{2}(b(\hat{v}_{1})+\hat{v}_{1})=\pi'(\hat{v}_{1})$.
Since $\hat{v}_{1}\ge r$, we have $b(\hat{v}_{1})\le b$,
which implies $\pi(\hat{v}_{1},b,r;\underline{v}_{1})=F_{2}(b+r)(\hat{v}_{1}-b)+(1-F_2(b+r))r\le F_{2}(b(\hat{v}_{1})+\hat{v}_{1})(\hat{v}_{1}-b(\hat{v}_{1}))+(1-F_{2}(b(\hat{v}_{1})+\hat{v}_{1})))\hat{v}_{1}=\pi(\hat v_1)$. Consequently, for all $v_{1}$,
$\pi(v_{1},b,r;\underline{v}_{1})\le\pi(v_{1})$.

In summary, with the belief that bidder 1's type is $\underline{v}_{1}$,
it is not profitable for any type $v_{1}$ to deviate to any off-path
proposals. This completes the proof.

\section{Proof of Lemma \ref{best-resp-2}}\label{proof-best-resp-2}
\begin{proof}
Observe that bidder 2 never accepts $r$ if her type $v_2\le b+r$   and never accepts $b$ otherwise. That is, for any type $v_2\le b+r$, her best response is either to accept $b$ or reject the proposal ( and thus compete with bidder 1 in the auction). Similarly, for any type $v_2>b+r$, her best response is either to accept $r$ or reject the proposal. 

  As shown in ES, when bidder 2's decision-making problem is to determine whether to accept or reject a bribe $b$ (and thus compete non-cooperatively with bidder 1 in the auction), the decision rule requires that for any type $v_2$ and
  $v_2^b$, if type $v_2^b$ accepts $b$, then any type $v_2<v_2^b$ accepts $b$.

We show below that when bidder 2's decision-making problem is to determine whether to accept or reject a request $r$ (and thus compete non-cooperatively with bidder 1 in the auction), the decision rule requires that for any type $v_2$ and $ v_2'$, if type $v_2'$ accepts $r$, then any type $v_2>v_2'$ accepts $r$. 

First we suppose $b+r<\bar v_2$.
Let the set of types of bidder 1 sending proposal $(b,r)$  be $Q_{b,r}$. Suppose $v_2,v_2'\ge b+r$ and  $v_2>v_2'$. For bidder 2 with type $v_2$, the  difference between the expected payoffs from accepting $r$ and rejecting the proposal is $\Delta _{v_2} \equiv v_2-r-E[(v_2-v_1)
\mathbf{1}_{v_1<v_2}|v_1\in Q_{b,r}]$, where $
\mathbf{1}_{X}$ is the indicator function for event $X$.
Thus,
\begin{align*}
\Delta_{v_2}-\Delta_{v_2'}=&\ v_2-v_2'-(E[(v_2-v_1)\mathbf{1}_{v_1<v_2}|v_1\in Q_{b,r}]-E[(v_2'-v_1)\mathbf{1}_{v_1<v_2'}|v_1\in Q_{b,r}] ) \\
\ge &\  v_2-v_2'- (E[(v_2-v_1)\mathbf{1}_{v_1<v_2}|v_1\in Q_{b,r}] - 
E[(v_2'-v_1)\mathbf{1}_{v_1<v_2}|v_1\in Q_{b,r}]) \\
=&\  v_2-v_2'-E[(v_2-v_2')\mathbf{1}_{v_1<v_2}|v_1\in Q_{b,r}] \\
=&\   (v_2-v_2')(1-E[\mathbf{1}_{v_1<v_2}|v_1\in Q_{b,r}]) \\
\ge &\  0.
\end{align*}
Therefore, there exists a $v_2^r\ge b+r$ such that any type $v_2\ge v_2^r$ accepts $r$ and any type  $v_2\in(b+r,v_2^r)$ rejects the proposal. 

It also follows from above  that if $b+r\ge \bar v_2$,
then the statement in the lemma is automatically true.
This completes the proof.  
\end{proof}

\section{Proofs of the Facts}\label{prof-facts}

\subsection*{Proof of Fact \ref{convexPi}}

The fact $\pi'(v_1)=F_2(b(v_1)+v_1)$ can be obtained directly from the envelope
theorem. It follows directly from \eqref{ic-general} that  $\pi(v_1)=v_1-F_2(b(v_1)+v_1)b(v_1)$ and
from \eqref{ic-r-eq-v} that $b(v_1)+v_1$ is
strictly increasing.

In either the case of $\underline v_1<\underline v_2$ or $\underline v_1\ge\underline v_2$, type $v_1$ requests his own valuation, which is accepted
by bidder 2 with probability one and thus $\pi_1(\underline  v_1)=\underline v_1$.

 \subsection*{Proof of Fact \ref{aboveV2b}}
 
The derivative of $\pi(v_1,b,r,v_2^b,v_2^r)$ w.r.t. $v_1$ is
\begin{align}\label{pi-v'}
\pi_{v_1}'(v_1,b,r,v_2^b,v_2^r)=
\begin{cases}
F_2(v_2^b)   &\text{ if } v_1\le v_2^b, \\
F_2(v_1)   &\text{ if } v_1\in(v_2^b,v_2^r],\\
F_2(v_2^r)  &\text{ if } v_1>v_2^r.
\end{cases}
\end{align}   
Recall that $\pi'(v_1)=F_2(b(v_1)+v_1)$. Because $b(v_1)>0$ for any type $v_1>\underline v_1$,
\eqref{pi-v'} implies that given any $v_2^b$ and $v_2^r$, 
$\pi_{v_1}'(v_1,b,r,v_2^b,v_2^r)< \pi'(v_1)$ for any
type $v_1\ge v_2^b$. This completes the proof.

 \subsection*{Proof of Fact \ref{rAboveV}}
Recall that when bidder 2  receives a proposal $(b,r)$, accepting $r$ gives a payoff $v_2-r$, while rejecting the proposal gives a payoff $v_2-v_1$. Thus, if $r\le \underline v_1$, then the proposal is never rejected.

 Suppose $b+r<\bar v_2$. Then for type $v_2<b+r$, it is more profitable to accept $b$ than $r$. Hence, if $r\le \underline v_1$, bidder 2's best response is to accept $b$ if $v_2\le b+r$ and accept $r$ if $v_2>b+r$, i.e., $v_2^b=b+r=v_2^r$. So the expected payoff of type $v_1$ is $\pi(v_1,b,r,b+r,b+r)=F_2(b+r)(v_1-(b+r))+r$. Then
\begin{align*}
\pi(v_1)-\pi(v_1,b,r,b+r,b+r)&=v_1-F_2(b(v_1)+v_1)b(v_1)
-[F_2(b+r)(v_1-(b+r))+r] \\
&= (1-F_2(b+r))(v_1-r)+F_2(b+r)b-F_2(b(v_1)+v_1)b(v_1).
\end{align*}
 If $b(v_1)<b$ and $b(v_1)+v_1<b+r$, then obviously $\pi(v_1)-\pi(v_1,b,r,b+r,b+r)>0$.
Because $r\le \underline v_1$, if  $b(v_1)<b$ and $b(v_1)+v_1\ge b+r$, or  $b(v_1)\ge b$, then  $\pi_{v_1}'(v_1,b,r,b+r,b+r)=F_2(b+r)\le F_2(b(v_1)+v_1)=\pi'(v_1)$. Since $r\le \underline v_1$, we have that
\begin{align*}
\pi(\underline v_1,b,r,b+r,b+r)&=F_2(b+r)(\underline v_1-b)+(1-F_2(b+r))r \\
&\le  F_2(b+r)\underline v_1 +(1-F_2(b+r))r \\
&\le \underline v_1=\pi(\underline v_1).
\end{align*}
Hence,  $\pi(v_1)\ge \pi(v_1,b,r,b+r,b+r)$ for any $v_1$  if $r\le \underline v_1$. 

Suppose $b+r\ge \bar v_2$. Then $v_2^b=v_2^r=\bar v_2$.
Then $\pi(v_1,b,r,\bar v_2,\bar v_2)=v_1-b$. So
\begin{align*}
\pi (v_1)-\pi(v_1,b,r,\bar v_2,\bar v_2)=b-F_2(b(v_1)+v_1)b(v_1).
\end{align*}
 If $b\ge b(v_1)$, then obviously $\pi (v_1)-\pi(v_1,b,r,\bar v_2,\bar v_2)\ge0$. If $b<b(v_1)$, then since $r\le
\underline  v_1$, we have $b(v_1)+v_1\ge b+r\ge \bar v_2$,
which implies $\pi'(v_1)=F_2(b(v_1)+v_1)=1=\pi_{v_1}'(v_1,b,r,\bar v_2,\bar v_2)$. Again, since $\pi(\underline v_1)\ge
\pi(\underline v_1,b,r,\bar v_2,\bar v_2)$, we have
$\pi(  v_1)\ge
\pi(v_1,b,r,\bar v_2,\bar v_2)$ for any
$v_1$.   
  
  \subsection*{Proof of Fact  \ref{bAboveV}}
If  $r>\underline v_1$ and $b\ge \bar v_2- \underline v_1$, then the bribe is accepted by all types of bidder 2, since $b\ge v_2-\underline v_1$ implies that for any type $v_2$, accepting $b$ is more profitable than competing with bidder 1 in the auction, and  $b\ge \bar v_2- \underline v_1>v_2-r$ implies that accepting $b$ is more profitable than accepting $r$--i.e., the best response of bidder 2 is described by $v_2^b=\bar v_2=v_2^r$. So the expected payoff of bidder 1 with type $v_1$  is $\pi (v_1,b,r,\bar v_2,\bar v_2)=v_1-b$. If $\bar v_1< \bar v_2$, then $\pi(\bar v_1,b,r,\bar v_2,\bar v_2)=\bar v_1-b<\bar v_2-b\le \underline v_1= \pi(\underline v_1)$. Since $\pi(v_1,b,r,\bar v_2,\bar v_2)$ is non-decreasing in $v_1$ and $\pi(v_1)$ is strictly increasing, $\pi(v_1,b,r,\bar v_2,\bar v_2)<\pi(v_1)$ for any type  $v_1\in[\underline v_1,\bar v_1]$.   If $\bar v_1\ge \bar v_2$,  then for  $v_1=\bar v_2$, $\pi (\bar v_2,b,r,\bar v_2,\bar v_2)=\bar v_2-b\le \underline v_1=\pi(\underline v_1)$. By the monotonicity of  $\pi(v_1,b,r,\bar v_2,\bar v_2)$ and $\pi(v_1)$, we have $\pi(v_1,b,r,\bar v_2,\bar v_2)\le \pi(v_1)$ for any type $v_1\le \bar v_2$. Then by Fact \ref{aboveV2b}, $\pi(v_1,b,r,\bar v_2,\bar v_2)\le \pi(v_1)$ for any type $v_1\in[\underline v_1,\bar v_1]$.      

\subsection*{Proof of Fact \ref{piBelow}}
From \eqref{payoff-off}, 
\begin{equation*}
\pi(v_1,b,r,b+\underline v_1,\bar v_2)=F_2(b+\underline v_1)(v_1-b)+\int_{\min\{b+\underline v_1,v_1\}}^{v_1} (v_1-v_2)f_2(v_2) dv_2.
\end{equation*}

If $b>0$, then $b+\underline v_1>\underline v_1$. Since $\pi(v_1,b,r,b+\underline v_1,\bar v_2)$ is non-decreasing and $\pi(v_1)$ is  strictly increasing, if $\bar v_1\le b+\underline v_1$, then  $\pi(\bar v_1,b,r,b+\underline v_1,\bar v_2)=F_2(b+\underline v_1)(\bar v_1-b)\le \underline v_1=\pi(\underline v_1)$,  which in turn implies  that $\pi(v_1,b,r,b+\underline v_1,\bar v_2)< \pi(v_1)$ for any type  $v_1 \in[\underline v_1,\bar v_1]$. If   $\bar v_1> b+\underline v_1$, then $\pi(b+\underline v_1,b,r,b+\underline v_1,\bar v_2)=F_2(b+\underline v_1)\underline v_1\le  \underline v_1=\pi(\underline v_1)$. By the  monotonicity of $\pi(v_1,b,r,b+\underline v_1,\bar v_2)$ and $ \pi(v_1)$, we have $\pi(v_1,b,r,b+\underline v_1,\bar v_2)<\pi(v_1)$  for any type $v_1\le b+\underline v_1$. Then by Fact \ref{aboveV2b}, $\pi(v_1,b,r,b+\underline v_1,\bar v_2)< \pi(v_1)$ for any type $v_1\in[\underline v_1,\bar v_1]$.  

If $b=0$ and $\underline v_1>0$, then $\pi(b+\underline v_1,b,r,b+\underline v_1,\bar v_2)=F_2(\underline v_1)\underline v_1< \underline v_1=\pi(\underline v_1)$ since we assume $\underline v_1<\bar v_2$. By the similar arguments in the case of $b>0$, we also have $\pi(v_1,b,r,b+\underline v_1,\bar v_2)< \pi(v_1)$ for any type $v_1\in[\underline v_1,\bar v_1]$. This completes the proof.  \qed\bigskip

\section{Proof of Proposition \ref{robust}}\label{robust-prof}
By Fact \ref{rAboveV} and Fact  \ref{bAboveV}, we can focus on off-path proposals $(b,r) $ with $r>\underline v_1$ and $b<\bar v_2- \underline v_1$. Thus $b+\underline v_1<b+r$ and $b+\underline v_1<\bar v_2$. For convenience of exposition, below we consider only the case of $b+r\le \bar v_2$. The analysis of the case of $b+r>\bar v_2$ is identical and omitted.\footnote{See footnote \ref{footnote}.} 

The lowest belief of bidder 2 is  $v_1=\underline v_1$. If bidder 2  believes $v_1=\underline v_1$, accepting $b$ gives $b$ and rejecting the proposal gives $v_2-\underline v_1$. So any type $v_2\le b+\underline v_1<b+r$ always accepts  $b$, i.e., $v_2^b\ge b+\underline  v_1$. Hence,
below we can focus on the beliefs such that $v_2^b\in[b+\underline v_1,b+r]$ and $v_2^r\in[b+r,\bar v_2]$.\footnote{When $b+r>\bar v_2$, only the bribe will be   seriously considered by bidder 2. Thus, in this case, $v_2^b\in[b+\underline v_1,\bar v_2]$ and $v_2^r=\bar v_2$. So the analysis for this case is the direct translation from the current one--i.e., replacing $b+r$ and $v_2^r$ by $\bar v_2$ appropriately.\label{footnote}} 

Suppose $b=0$ and $\underline v_1=0$. When $v_2^b= \underline v_1=0$ and $v_2^r= \bar v_2$--i.e., the proposal is rejected by all types of bidder 2 and both bidders compete in the auction--only type $\underline v_1$ is indifferent between the equilibrium payoff and the payoff from the deviation, while all other types of bidder 1 are strictly worse off. To see this, the expected payoff from the deviation is the one from a standard second-price auction, which has a slope $F_2(v_1)$, which in turn is strictly smaller than $\pi'(v_1)=F_2(b(v_1)+v_1)$.   
 Hence, type $\underline v_1=0$ cannot be excluded by the D1 criterion,  and a reasonable belief of bidder 2 is that bidder 1's type is $v_1=\underline v_1=0$. Since $b=0$, the best response of bidder 2 with this belief is then to reject the proposal and compete with bidder 1 in the auction. Then the above confirms that it is not profitable for any type of bidder 1 to deviate to the off-path $(b=0,r)$. Hence, below we only need to examine the event $b>0$ or $\underline v_1>0$.

For a given pair of $(b,r)$, let 
\begin{equation}\label{m:f}
M(v_2^b,v_2^r)=\max_{v_1}\; \pi(v_1)-\pi(v_1,b,r,v_2^b,v_2^r).
\end{equation}
Clearly, for any given $v_2^b\in[b+\underline v_1,b+r]$ and $v_2^r\in[b+r,\bar v_2]$, $\pi(v_1)-\pi(v_1,b,r,v_2^b,v_2^r)$ is well defined on $[\underline v_1,\bar v_1]$ and
 continuous in $v_1$. By the extreme value theorem, it  admits a maximum for any $v_2^b,v_2^r$ in the relevant intervals. The solution to the maximization problem in \eqref{m:f} is continuous and thus $M(v_2^b,v_2^r)$ is continuous. 

Suppose that for a given proposal $(b,r)$, there exists no $v_2^b\in[b+\underline v_1,b+r]$ and no $v_2^r\in[b+r,\bar v_2]$ such  that $M(v_2^b,v_2^r)>0$. Then  $\pi(v_1,b,r,v_2^{b},v_2^{r})\le \pi(v_1)$ for any type $v_1$--i.e., it is not profitable for any type of bidder 1 to deviate to $(b,r)$ for any belief of bidder 2, and we are done.

Suppose for a given proposal $(b,r)$,  for some $v_2^b=v_2^{b'}\in[b+\underline v_1,b+r]$ and $v_2^r=v_2^{r'}\in[b+r,\bar v_2]$,   $M(v_2^{b'},v_2^{r'})>0$. Observe that  Fact \ref{piBelow} implies $M(b+\underline v_1,\bar v_2)<0$. By the intermediate value theorem, there exists some $v_2^b=v_2^{b*}\in[b+\underline v_1,v_2^{b'}]$ and $v_2^r=v_2^{r*}\in[ v_2^{r'}, \bar v_2]$ such that $M(v_2^{b*},v_2^{r*})=0$. Let $\mathbb{V}_1$ be the set of $v_1$ such that $M(v_2^{b*},v_2^{r*})=0$. Then by the fact that $\pi'(v_1)-\pi_{v_1}'(v_1,b,r,v_2^b,v_2^r)>0$ for all $v_1\ge v_2^{b}$ (from \eqref{pi-v'} and Fact \ref{convexPi}),   we can conclude that for any $v_1=v_*\in \mathbb{V}_1$,
\begin{equation}
v_*<v_2^{b*}\le b+r.
\end{equation}
By the fact that $\pi_{v_1}'(v_1,b,r,v_2^b,v_2^r)=v_2^b$ for all $v_1\le v_2^b$ and $\pi(v_1)$ is strictly convex (again from \eqref{pi-v'} and Fact \ref{convexPi}),  $\mathbb{V}_1$ is a singleton.

Thus, if for some type $v_1$ there exists some $v_2^b\in[b+\underline v_1,b+r]$ and $v_2^r\in[b+r,\bar v_2]$ such that $\pi(v_1,b,r,v_2^{b},v_2^{r})\ge \pi(v_1)$, then there must exist a  pair of $(v_2^{b*},v_2^{r*})$ such that $\pi(v_1,b,r,v_2^{b*},v_2^{r*})$ is \textit{tangent to}  $\pi(v_1)$ at some unique $v_1=v_*$, in the sense that for all $v_1\neq v_*$, $\pi(v_1,b,r,v_2^{b*},v_2^{r*})< \pi(v_1)$ and $\pi(v_*,b,r,v_2^{b*},v_2^{r*})= \pi(v_*)$. 
 Hence, type $v_*$ is not excluded by the D1 criterion. So  a reasonable belief of bidder 2 is that bidder 1's type is $v_1=v_*$ and we adopt this belief for the analysis below.

Suppose the tangency point is at $\underline v_1$ and thus a reasonable belief of bidder 2 is  $v_1=\underline v_1$. Then for bidder 2, accepting $b$ gives $b$, accepting $r$ gives $v_2-r$, and rejecting the proposal gives $v_2-\underline v_1$.  Since $r>\underline v_1$ and $b<\bar v_2- \underline v_1$, it is optimal for bidder 2 to accept $b$ if $v_2\le b+\underline v_1$ and reject the proposal otherwise. Then Fact \ref{piBelow} implies that it is not profitable for any type of bidder 1 to deviate to the proposal. 

It follows, then, that  below we can focus on interior tangency points, i.e, $v_*\in(\underline v_1,v_2^{b*})$. It in turn follows that 
$\pi_{v_1}'(v_*,b,r,v_2^b,v_2^r)=\pi'(v_*)$, which implies  $v_2^{b*}=b(v_*)+v_*$.   

Suppose $v_2^{b*}=b+r=v_2^{r*}$, which implies  $b(v_*)+v_*=b+r$. If $v_*\ge r$, then the best response of bidder 2 is to accept $b$ if $v_2\le b+r$ and accept $r$ otherwise. Then the expected payoff of any type $v_1$ from the best response is  exactly $\pi(v_1,b,r,v_2^{b*},v_2^{r*})$, which, by tangency, is no greater than $ \pi(v_1)$    and we are done. The sub-case $v_*<r$  is analyzed   below together with the case of    $v_2^{b*}<b+r$ or $b+r<v_2^{r*}$.

Suppose   $v_2^{b*}<b+r$ or $b+r<v_2^{r*}$. Then we can always conclude $v_*<r$. To see this, observe that  type   $v_*$ is indifferent between  $\pi(v_*,b,r,v_2^{b*},v_2^{r*}) $ and $\pi(v_*)$, i.e.,
\begin{equation*}
F_2(v_2^{b*})(v_*-b)+(1-F_2(v_2^{r*}))r=v_*-F_2(b(v_*)+v_*)b(v_*),
\end{equation*}
which, by the fact that $v_2^{b*}=b(v_*)+v_*$, can be rearranged into
\begin{equation*}
v_*-r=F_2(v_2^{b*})(b(v_*)+v_*-b)-F_2(v_2^{r*})r.
\end{equation*} 
Thus, if $v_2^{b*}<b+r$ or  if $b+r< v_2^{r*}$, then
$v_2^{b*}<v_2^{r*}$ and $v_2^{b*}=b(v_*)+v_*\le b+r$, which in
turn implies $v_*<r$. 

Therefore, below we can focus on the belief $v_1=v_*<r$,  with
which
the best response of bidder 2 is to accept $b$ if $v_2\le b+v_*$ and reject the proposal otherwise. The
expected payoff of bidder 1 from the proposal is then $\pi(v_1,b,r,b+v_*,\bar v_2)$. 

By Fact \ref{aboveV2b}, we can for now restrict our attention to type $v_1\le b+v_*$. For any type $v_1\le b+v_*$, the expected payoff is 
\begin{equation*}
\pi(v_1,b,r,b+v_*,\bar v_2)=F_2(b+v_*)(v_1-b).
\end{equation*}
The tangency condition says that for any type $v_1$, 
\begin{align*}
\pi(v_1,b,r,v_2^{b*},v_2^{r*})&=F_2(b(v_*)+v_*)(v_1-b)+(1-F_2(v_2^{r*}))r \\
&\le v_1-F_2(b(v_1)+v_1)b(v_1)  \\
&=   \pi(v_1).
\end{align*}

Suppose $b\le b(v_*)$. Then clearly $\pi(v_1,b,r,b+v_*,\bar v_2)\le \pi(v_1,b,r,v_2^{b*},v_2^{r*})\le \pi(v_1)$ for any type  $v_1$.

Suppose $b> b(v_*)$. Then
\begin{align*}
\pi(v_*)-\pi(v_*,b,r,b+v_*,\bar v_2) 
&=v_*-F_2(b(v_*)+v_*)-F_2(b+v_*)(v_*-b) \\
&=v_*(1-F_2(b+v_*))+F_2(b+v_*)b-F_2(b(v_*)+v_*)b(v_*)\\
&\ge 0.
\end{align*}
 Furthermore, recall that $\pi(v_1)$ is strictly convex
 and observe that $\pi(v_*,b,r,b+v_*,\bar v_2)$ has
 a constant slope for  $v_1<v_*$.
 Thus  $b+v_*>b(v_*)+v_*$  implies  $\pi_{v_1}'(v_1,b,r,b+v_*,\bar v_2)=F_2(b+v_*)>F_2(b(v_1)+v_1)=\pi'(v_1)$ for any type $v_1<v_*$. So for any type $v_1<v_*$, $\pi(v_1,b,r,b+v_*,\bar v_2)< \pi(v_1)$. For any type $v_1\in(v_*,b+v_*]$, 
\begin{equation*}
\pi(v_1)-\pi(v_1,b,r,b+v_*,\bar v_2)=v_1(1-F_2(b+v_*)) +F_2(b+v_*)b -F_2(b(v_1)+v_1)b(v_1).
\end{equation*}
If $b(v_1)<b$ and $b(v_1)+v_1<b+v_*$, then obviously $\pi(v_1)-\pi(v_1,b,r,b+v_*,\bar v_2)>0$. If $b(v_1)<b$ and $b(v_1)+v_1\ge b+v_*$, or if $b(v_1)\ge b$, then 
$\pi'(v_1)=F_2(b(v_1)+v_1)>F_2(b+v_*)=\pi_{v_1}'(v_1,b,r,b+v_*,\bar v_2)$ for any type $v_1\in(v_*,b+v_*]$. Since,  from the above, $\pi(v_*)>\pi(v_*,b,r,b+v_*,\bar v_2)$, we have that for any $v_1\in(v_*,b+v_*]$, $\pi(v_1)>\pi(v_1,b,r,b+v_*,\bar v_2)$. Fact \ref{aboveV2b} then implies $\pi(v_1)>\pi(v_1,b,r,b+v_*,\bar v_2)$ for any $v_1 \in[\underline v_1,\bar v_1]$.

 Aggregating all of the above cases, the proof of the equilibrium that
 survives the D1 criterion is completed.

\section{Proof of Proposition \ref{sep-pi-vs-es} }\label{prof-sep-pi-vs-es}

In our model the expected payoff of bidder 1 is $\pi(v_1)=v_1-F_2(b(v_1)+v_1)b(v_1)$ and  $\pi'(v_1)=F_2(b(v_1)+v_1)$. In a robust equilibrium with a bribing function $B(v_1)$ in ES, the expected payoff of bidder 1 is $\Pi(v_1)=F_2(B (v_1)+v_1)(v_1-B (v_1))$ and the envelope theorem implies that $\Pi'(v_1)=F_2( B(v_1)+v_1)$ for type $v_1\le \hat v_1$, which is also true for type $v_1> \hat v_1$ in fact. Thus
\begin{equation*}
\pi(v_1)-\Pi(v_1)=v_1(1-F_2(B(v_1)+v_1))+F_2(B(v_1)+v_1)B(v_1)-F_2(b(v_1)+v_1)b(v_1).
\end{equation*}
Thus, whenever $B(v_1)\ge b(v_1)$, $\pi(v_1)-\Pi(v_1)\ge 0$. On the other hand, whenever $B(v_1)<b(v_1)$, $\pi'(v_1)>\Pi'(v_1)$. Since $\pi(\underline v_1)=\underline v_1\ge \Pi(\underline v_1)$, $\pi(v_1)\ge \Pi(v_1)$ for any type $v_1$.
 
\section{Proof of Proposition \ref{other-sep}} \label{prof-other-sep}

Suppose first that $\gamma(v_1)\le v_1$ for any type $v_1$. Clearly,  $\pi(v_1;\beta,\gamma)=F_2(\beta(v_1)+\gamma(v_1))(v_1-(\beta(v_1)+\gamma(v_1)))+\gamma(v_1)$. The envelope theorem implies that $\pi'(v_1;\beta,\gamma)=F_2(\beta(v_1)+\gamma(v_1))$. Thus,
\begin{align*}
\pi(v_1)-\pi(v_1;\beta,\gamma)&=v_1-F_2(b(v_1)+v_1)b(v_1)\\&\quad-[F_2(\beta(v_1)+\gamma(v_1))(v_1-(\beta(v_1)+\gamma(v_1)))+\gamma(v_1)] \\
&= (v_1-\gamma(v_1))(1-F_2(\beta(v_1)+\gamma(v_1)))\\&\quad+F_2(\beta(v_1)+\gamma(v_1))\beta(v_1) -F_2(b(v_1)+v_1)b(v_1).
\end{align*}
 If $b(v_1)<\beta(v_1)$ and $b(v_1)+v_1<\beta(v_1)+\gamma(v_1)$, then obviously $\pi(v_1)-\pi(v_1;\beta,\gamma)>0$.  If $b(v_1)<\beta(v_1)$ and $b(v_1)+v_1\ge \beta(v_1)+\gamma(v_1)$, or if $b(v_1)\ge \beta(v_1)$, then $\pi'(v_1)=F_2(b(v_1)+v_1)\ge F_2(\beta(v_1)+\gamma(v_1))=\pi'(v_1;\beta,\gamma)$. Since $\pi(\underline v_1)=\underline v_1\ge \pi(\underline v_1;\beta,\gamma)$, $\pi(v_1)\ge\pi(v_1;\beta,\gamma)$ for any type $v_1$.

Suppose now that $\gamma(v_1)> v_1$ for some type $v_1$. For any such type $v_1$, the request is never accepted and the bribing function $\beta(v_1)$ must satisfy the incentive compatibility condition in ES's model. Consequently, the expected payoff of any such type $v_1$ must be no greater than the one in the equilibrium with $r(v_1)=v_1$. This completes the proof.

\section{Proof of Proposition \ref{nonexistence-1st} }\label{prof-nonexistence-1st} 

Suppose there exists an arbitrary separating equilibrium in which $r(v_1)\le v_1$, and thus the proposals are accepted with certainty. In particular, if $b(v_1)+r(v_1)<\bar v_2$, then any type $v_2 \ge b(v_1)+r(v_1)$ accepts the request. In the equilibrium, for type $v_{1}=0$, $b(0)=r(0)=0$, and he earns zero payoff. Clearly, for any type  $v_1>0$, if $b(v_1)=0$, then $r(v_1)>0$; otherwise type $v_1$ is pooling with type zero. So in any such separating equilibrium,   if there exists a type $v_1>0$ offering $b(v_1)=0$, then $r(v_1)= \tilde r$ for some $\tilde r >0$. Let the set of such $\tilde r$ be $\tilde R$. Suppose $\tilde R$ is not empty.  Suppose $\inf\{\tilde R \}=0$. Then obviously there exists some on-path proposal with $b(v_1)=0$ and  $r(v_1)<\bar v_2$. Then any type $v_2 >r(v_1)$ accepts $r(v_1)$--i.e., a positive measure of type $v_2$ accepts $r(v_1)$. But then it is profitable for type zero to deviate to this on-path proposal and is thus a contradiction.   So if  $\tilde R$ is not empty, then in any such separating equilibrium with $r(v_1)\le v_1$,  there exists some $\hat r=\inf\{\tilde R \}>0$ such that proposals $\left(b=0,r\in(0, \hat r)\right) $ are off the path. On the other hand, if $\tilde R$ is empty, then any proposal with zero bribe is off the path. 

Consider an off-path proposal $\left(b=0,r>0\right)$ from some small $r$. In order for
type $v_{1}=0$ to not benefit from the proposal, bidder 2's best
response must involve no acceptance of request $r$. That is, some
types of bidder 2 accept $b=0$, and the other types must reject the
proposal (except possibly for a zero-measure  of types accepting $b$). If bidder 1 meets  a type $v_2$ accepting $b=0$, then  any type $v_1>0$ realizes his valuation.  Now suppose there is a positive measure (e.g., denoted by $P_{2r}$) of types that
reject the proposal and both bidders are in the continuation game--i.e., the first-price auction. In the auction,
the highest bid of bidder 2 must not exceed $r$, since otherwise
it is more profitable to accept $r$ in the first place. Therefore,
the highest type of bidder 1 can at least bid $r+\varepsilon$ and
win the object and the payoff is $\bar{v}_{1}-(r+\varepsilon)$. So the expected payoff from deviating to the off-path proposal is at least $(1-P_{2r})\bar v_1+P_{2r}[\bar{v}_{1}-(r+\varepsilon)]$, the minimum of which is $\bar{v}_{1}-(r+\varepsilon)$. In the separating equilibrium, the expected payoff of type $\bar v_1$ is $\pi_{1}(\bar{v}_{1})=F_{2}(b(\bar{v}_{1})+r(\bar{v}_{1}))(\bar{v}_{1}-b(\bar{v}_{1}))+\left(1-F_{2}(b(\bar{v}_{1})+r(\bar{v}_{1}))\right)r(\bar{v}_{1})$.
Since $b(\bar v_1 )>0$ and $r(\bar{v}_{1})\le \bar v_1$, $\pi_{1}(\bar{v}_{1})<\bar v_1$. Therefore, for low enough $r+\varepsilon$, we have $\bar{v}_{1}-(r+\varepsilon)>\pi_{1}(\bar{v}_{1})$--that is, the highest type of bidder 1 earns a higher payoff
from the off-path proposal than the equilibrium payoff. This completes the proof.

\bibliographystyle{TeNoURL}
\bibliography{bribing}

\end{document}